\documentclass[english,twocolumn,showpacs,preprintnumbers,amsmath,amssymb]{revtex4}
\usepackage{subfigure}
\usepackage{graphicx}
\usepackage{amssymb}

\makeatletter

\usepackage{babel}
\makeatother

\begin{document}

\title{Multifractal analysis of nonhyperbolic coupled map lattices:
Application to genomic sequences}

\author{A. Provata$^{1,2}$ and C. Beck$^{2}$}

\affiliation{$^1$Institute of Physical Chemistry, National Center for Scientific
Research ``Demokritos\char`\"{}, 15310 Athens, Greece \\
$^2$School of Mathematical Sciences, Queen Mary University of
London, Mile End Road, London E1 4NS,UK}

\begin{abstract}

Symbolic sequences generated by coupled map lattices (CMLs)
can be used to model
the chaotic-like structure of genomic sequences. In this study
it is shown that diffusively coupled Chebyshev maps of order 4
(corresponding to a shift of 4 symbols)
very closely reproduce the multifractal spectrum $D_q$ of human
genomic sequences for
coupling constant $\alpha =0.35\pm 0.01$ if $q>0$.
The presence of rare configurations
causes deviations for $q<0$, which disappear
if the rare event statistics of the CML is modified.
Such rare configurations are known
to play specific functional roles in genomic sequences serving as promoters
or regulatory elements.

\end{abstract}


\date{\today}

\maketitle

\section{Introduction}
\label{sec1}
Coupled Map Lattices (CMLs) are frequently used as models for complex,
often chaotic spatial and dynamical structures observed in diverse physical systems
\cite{kaneko, kapral, chate, mackey}.
Particular CMLs have been used to model hydrodynamic systems, chemical kinetics,
biological systems, and field theoretical models \cite{kanekobook, beckbook}.
These types of models arise normally in situations where the nonlinear
nature of the phenomena is complimented by a nontrivial underlying spatial geometry.
Of particular interest are non-hyperbolic
coupled map lattices, where the local map is allowed to have one or several points
with zero slope.

In the current study we investigate the behavior of
coupled 4-th order Chebyshev maps, $T_4$, and compare their multifractal spectra
with that of
DNA sequences. In fact, it is well-known that
the dynamics of $T_4$ is equivalent to a Bernoulli shift of 4 symbols.
Thus the coupled map dynamics with $T_4$ as a local map corresponds to a coupled shift of information
that is encoded by 4 symbols.
In this respect it is natural to study the potential correspondence of
statistics generated by coupled $T_4$ as compared to that of genomic sequences
composed of the four symbol-nucleotides, namely Adenine (A),
Cytosine (C), Guanine (G) and Thymine (T). Earlier studies on the
primary structure of DNA have shown that
the statistics of genomic sequences exhibits nontrivial correlations and cannot be
reproduced by a pure random stochastic process involving 4 symbols
\cite{li:1992,peng:1992,voss:1992,mantegna:1994,scaling,grosse:2002,carpena:2007,usatenko:2003,afreixo:2004,li:2002,bernaola:2004,li:2004,cheng:2005,messer:2006,katsaloulis:2002,katsaloulis:2006,han:2008,freudenberg:2009,beck:2010}.
A natural way to gradually introduce  correlations in the phase space structure of $T_4$ is
 via coupling of many $T_4$ maps on a lattice.
In our approach nontrivial correlations are introduced
by means of a coupling constant $\alpha$ which diffusively couples nearest neighbor
 maps on the lattice and takes values $0\le\alpha\le 1$.
\par 
Chebyshev maps are known to exhibit the strongest possible chaotic behavior
characterized by a minimum skeleton of higher-order correlations \cite{nonli, hilgers}.
For weak coupling $\alpha$ analytic results have been previously derived on the
perturbed invariant 1-point density
\cite{groote:2006} and on the existence of periodic orbits \cite{dettmann}.
These investigations provide motivation
for a discussion of the possibility of CMLs to reproduce
similar statistics as observed in genomic sequences for finite values of the coupling constant. In this study,
we investigate the multifractal spectrum resulting by appropriately
sectioning the phase space of the CML to assimilate 4-symbol sequences.
The choice of the multifractal spectrum 
as the relevant observable is particularly suitable for comparing genomic
and CML sequences because it reveals the characteristic details of moments
and symbol correlations of all orders.
\par In the next section we first recall the multifractal spectrum of a single Chebyshev map
and we further explore the spectra of coupled Chebyshev maps on a 1-D lattice
with periodic boundary conditions. In section
\ref{sec3} a 1-1 correspondence is introduced between 4 appropriately
chosen sections of the local CML phase space and the 4 symbols of an artificial genomic sequence.
 The multifractal spectra of entire human chromosomes are compared
with the CML spectra for various values of $\alpha$. Coupling values of the order of
$\alpha \sim 0.34-0.36$ are shown to yield multifractal spectra closely approximating the correlations in
DNA  sequences for positive $q$. In section \ref{sec4} it is shown 
that rare configurations need to be introduced
in the CML dynamics to closely approximate the DNA spectra for both positive and negative $q$.
In the concluding section the final results are summarized and open
problems are discussed.

\section{Multifractal Spectra of CMLs}
\label{sec2}
\subsection{Multifractal Spectrum of 4-th order Chebyshev map}
\label{sec2.1}
\begin{figure}
\includegraphics[clip,width=0.45\textwidth,angle=0]{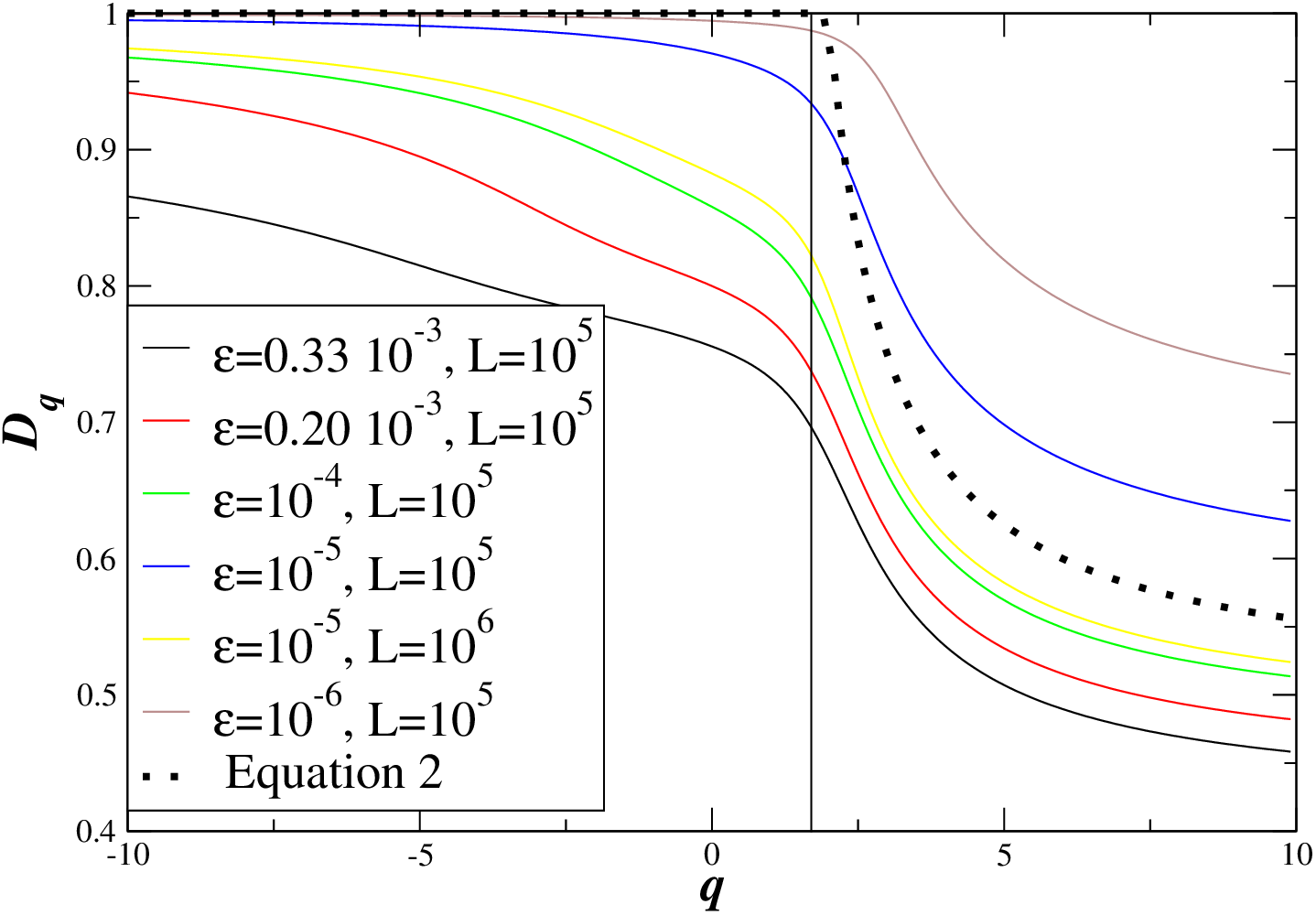}

\caption{\label{fig:01}{} (Color online)
The multifractal spectrum of uncoupled ($\alpha =0$) $T_4$'s
as obtained numerically for finite $\epsilon$ and $N$.
The solid vertical line is the line $q=2$.
The dotted curve represents the analytical expression of the multifractal
spectrum, Eq. (\ref{eq01}). }
\end{figure}

The dynamics of the $T_4$ map  is
generated by the recurrence relation
\begin{eqnarray}
\begin{array}{lll}
x_{n+1} & = & T_4(x_n)=8 x_n^4-8_n^2+1,\\
\\
      & {\rm with} & n=0,1,2..., \>\> -1\le x_0\le 1.
\end{array}
\label{eq0-1}
\end{eqnarray}
$x_n\in [-1,1]$ is a continuous variable and takes values in the interval $[-1,1]$
and $n$ is a discrete time variable. This map is known to show strongest possible
chaotic behavior. The multifractal spectrum generated by its invariant density
is known analytically to take the form \cite{jstatphys,schloeglbook}:
\begin{eqnarray}
D_q=\left\{
\begin{array}{lll}
1, \>\> &{\rm for}& \>\> q \le 2 \\
\frac{q}{q-1}\frac{1}{2}, \>\> &{\rm for}& \>\> q > 2
\end{array}
\right.
\label{eq01}
\end{eqnarray}
Generally the Renyi (multifractal) dimensions are defined as
\begin{equation}
D_q =\lim_{\epsilon \to 0} \frac{1}{q-1} \frac{1}{\log \epsilon} \log \sum_i p_i^q
\end{equation}
where $p_i= \int_{i-th\; box} p(x)dx$ are the probabilities associated with
a partitioning of the phase space into boxes of equal size $\epsilon$.
The multifractal spectrum given by Eq.~(\ref{eq01}) is easily obtained from the
invariant probability density
\begin{eqnarray}
p(x)=\frac{1}{\pi \sqrt{1-x^2}},\>\> -1\le x\le 1,
\label{eq02}
\end{eqnarray}
see, e.g. \cite{schloeglbook}.
The presence of two singularities of the probability density at $x=\pm 1$
 produces a phase transition-like point of $D_q$ at $q=2$,
see Eq. (\ref{eq01}) and the corresponding $D_q$ vs. $q$
diagram in Fig. \ref{fig:01}.
This multifractal spectrum
is formally obtained in the limit of infinitesimal ($\epsilon \to 0$)
segmentation of the
interval $[-1,1]$ where $T_4$ is defined. This idealized limit is hardly
observable in finite size systems, as is demonstrated in Fig. \ref{fig:01}.
In particular, genomic sequences
are finite in size, having a definite number of nucleotides,
hence it is not possible to achieve infinitesimally small segmentations.
For comparison with real data it is useful to
explore finite size effects, for small but nonzero values of
$\epsilon$.  These are depicted in Fig. \ref{fig:01}. For statistical reasons,
large numbers of $L=10^5$ or
$10^6$ uncoupled Chebyshev maps were considered, iterated over
5000 time steps with random initial values. The analytical result
(black dotted curve)
is obtained in the limit $\epsilon \to 0$ and $L\to \infty$.

For finite $\epsilon >0$ the abrupt
critical point behavior is deformed into a
smooth but rapidly changing curve
in the region of $0<q<2$. The second derivative of $D_q$
as a function of $q$ is sometimes
observed to switch sign. Such 'humps'
have also been observed for the
Renyi dimensions and Renyi entropies associated with symbol sequences of the
human genome, see Refs. \cite{yu:2001,provata:2010,beck:2010} for more details.

Generally, the shape of numerically determined $D_q$ spectra
of non-hyperbolic maps is
heavily influenced by finite size effects, and we expect similar finite size effects
to be present for multifractal spectra of genomic sequences.

\subsection{Multifractal Spectra of Linearly Coupled Chebyshev Maps }
\label{sec2.2}

Our interest in coupling Chebyshev maps results from the tendency of
local interactions in systems with multiple components. In the current
study we consider the simplest possible diffusive nearest neighbor coupling on
a 1-D lattice with periodic boundary conditions. Namely, we assume a
linear chain of units ('particles') each of which is labelled by the index $i$.
Each unit  evolves according to
Eq. (\ref{eq0-1}) with additional, equal contributions from the left and right
nearest neighbor particles.
In the linear chain of size $L$, a coupling $\alpha$ is introduced so
that the variable $x^i$ of the $i-th$ unit follows the
recurrence relation
\begin{eqnarray}
x_{n+1}^i=(1-\alpha)T_4(x_n^i)+\frac{\alpha}{2} \left(
T_4(x_{n+1}^i)+T_4(x_{n-1}^i)\right)
\label{eq03}
\end{eqnarray}
As initial conditions, a random distribution of $x_0^i \in \left[
-1,1\right]$ is assumed.
When Chebyshev maps are coupled on lattices the
invariant 1-point densities are gradually deformed and singularities tend to smooth out
\cite{groote:2006}.
In particular, for
$\alpha =1$ there are no singularities and $D_q=1, \forall q$,
while the case $\alpha =0$ corresponds to a collection of
independent $T_4$'s and the spectrum is given by the relation
(\ref{eq01}).
\par
\begin{figure}
\includegraphics[clip,width=0.45\textwidth,angle=0]{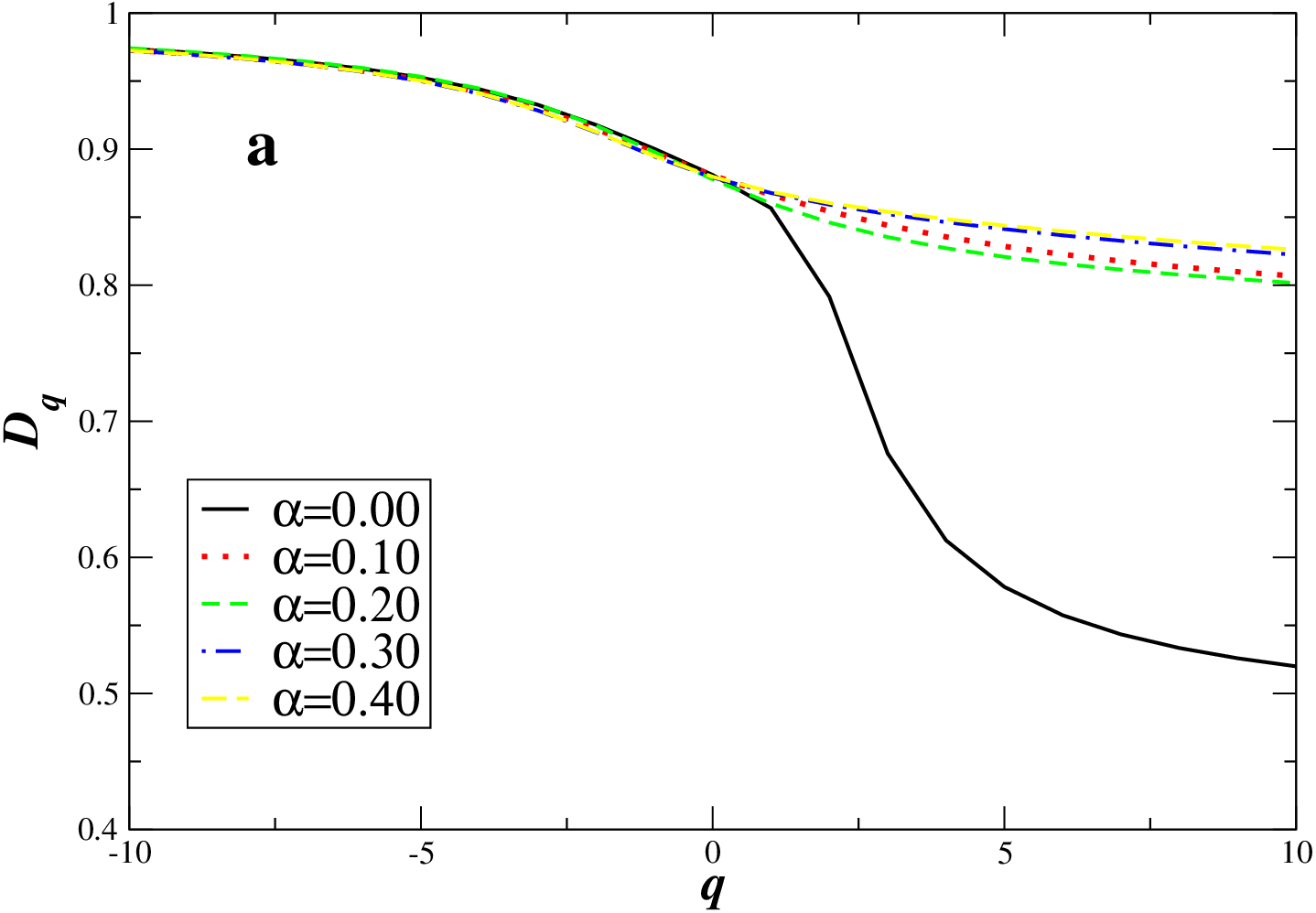}
\includegraphics[clip,width=0.45\textwidth,angle=0]{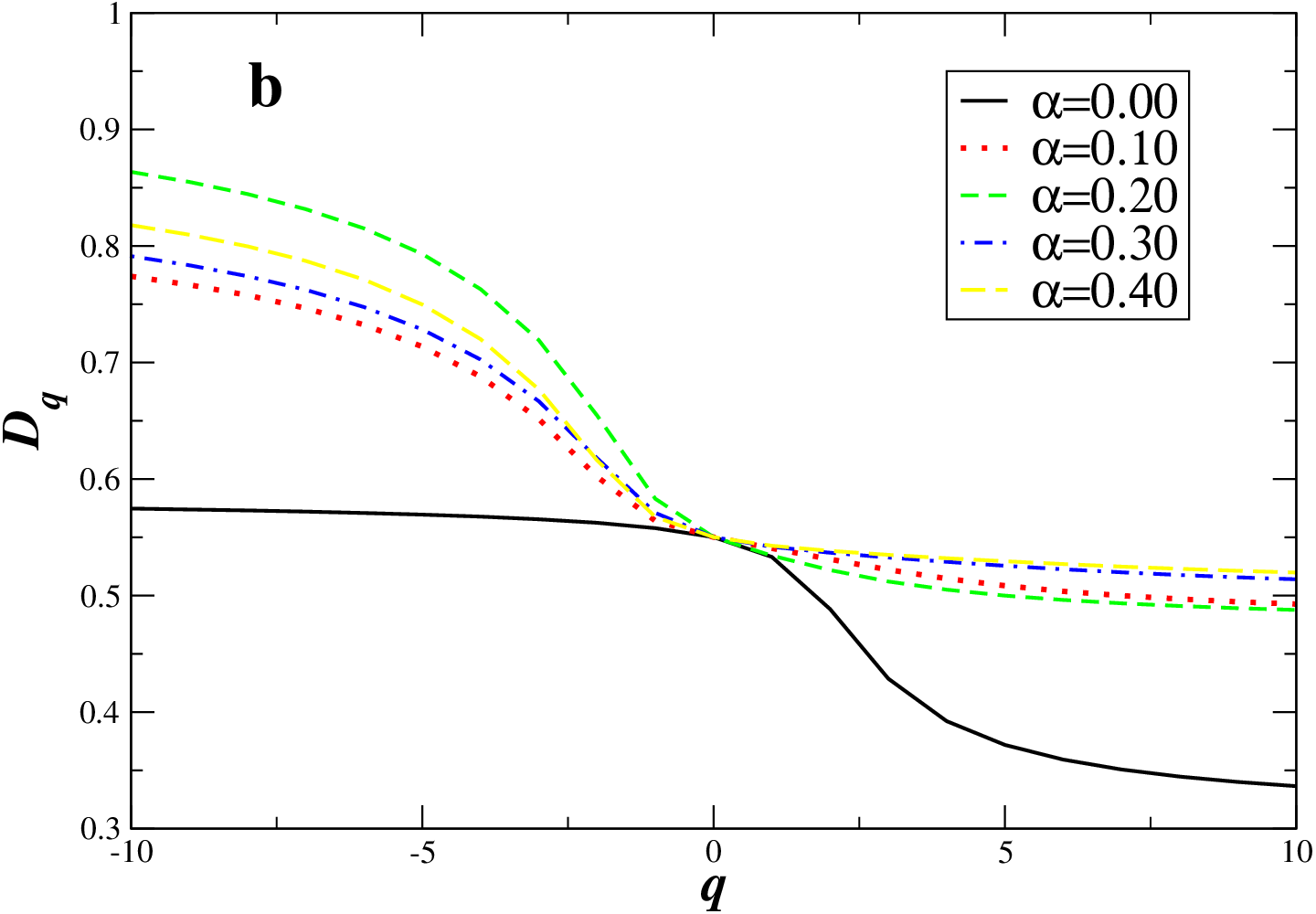}


\caption{\label{fig:02}{} (Color online)
 a) Numerically obtained multifractal spectra of coupled $T_4$'s for various values of
$\alpha$, $\epsilon =10^{-5}$. b)  Humps in the multifractal spectra
for coupled $T_4$ are observed for relatively large values of  $\epsilon$
(here $\epsilon =10^{-3}$ is shown). $L=10^6$ in
both a and b plots.}
\end{figure}

 In Fig. \ref{fig:02} a series of multifractal spectra
for different values of $\alpha$ are shown, for finite but small
values of $\epsilon$. $L=10^6$ coupled $T_4$ maps are
taken into account with $\epsilon = 10^{-5}$ and $\epsilon =10^{-3}$
in Figs. \ref{fig:02}a and \ref{fig:02}b, respectively. In both cases, as
$\alpha \to 1$, the multifractal spectrum
is seen to become uniform. In Fig. \ref{fig:02}b humps are observed due to
finite size effects ($\epsilon$
as large as $10^{-3}$), for different values of $\alpha$.
Note,  in Fig. \ref{fig:02}b, that the humps are more evident for higher
values of the coupling constant
$\alpha $. These $\alpha $-values are consistent with those that reproduce
similar statistics as DNA sequences, as will be seen in the subsequent sections.

\section{Symbolic Sequences Resulting from CMLs}
\label{sec3}
Having analyzed multifractal spectra
that are directly associated with the local
distribution of the state variables $x_n^i$ of the
CML we now want to go a step further and produce symbol
sequences from the CML. 
Comparing with the multifractal
spectra of genomic sequences
\cite{yu:2001,gutierrez:2001,su:2009,provata:2010} we note that
there are certain similarities in the two spectra which suggests
to explore the possibility of a certain chaotic CML processes to reproduce
the most important genomic spectral features.
\par As a first step in this direction one has to reconstruct a symbolic
sequence of 4 letters, based on the distribution of the local CML variable $x$.
When constructing the artificial symbolic sequence one needs, at least,
to respect the symbol concentrations of the original genomic sequence.
If the genomic populations (mean concentrations) of the 4 nucleotides are denoted as
$p_A,p_C,p_G,p_T=1-p_A-p_C-p_G$ for Adenine, Cytosine, Guanine and
Thymine, respectively, then the artificial $T_4$-based genomic sequence
should contain the same frequency of the symbols. To achieve consistency
between the genome basepair population and the symbolic sequence one
needs to consider again the 1-point distribution $p(x)$ of the
coupled $T_4$ map.
The interval $[-1,1]$ is segmented into four subintervals $[-1,x_1]$
,$[x_1,x_2]$,$[x_2,x_3]$ and $[x_3,1]$, to accommodate the 4 basepairs.
The values of $x_1,x_2$ and $x_3$ were chosen to fulfill the 
basepair frequency constraints:

\begin{eqnarray}
\begin{array}{lll}
\int_{-1}^{x_1}dx\;p(x)=p_A  \\
\int_{x_1}^{x_2}dx\;p(x)=p_C \\
\int_{x_2}^{x_3}dx\;p(x)=p_G   \\
\int_{x_3}^{1}dx\;p(x)=p_T   \\
\end{array}
\label{eq04}
\end{eqnarray}
Having fixed the segmentation values $[x_1,x_2,x_3]$,
the $i$-th symbol $S_n^i$ of the artificial genomic sequence is chosen as
\begin{eqnarray}
S_n^i=\left\{
\begin{array}{lll}
A  &{\rm if}& -1< x_n^i\le x_1 \\
C  &{\rm if}& x_1 < x_n^i \le x_2\\
G  &{\rm if}& x_2 < x_n^i \le x_3 \\
T  &{\rm if}& x_3 < x_n^i \le 1  \\
\end{array}
\right.
\label{eq05}
\end{eqnarray}
Hence, a symbolic sequence $S_n(i), i=1,...L$
is produced which on the one hand carries the complexity
of the CML and on the other hand respects the average concentrations of the
DNA sequence under consideration, $(p_A,p_C,p_G,p_T$).
\par In the search for a proper value of the coupling constant $\alpha$ which
best describes the complexity of the genomic sequences, it is important to
create CML-generated sequences of length comparable with the genomic ones.
In the current study sequences of $L=10^7$ were produced to assimilate
the chromosomal DNA.
To avoid transient phenomena and to ensure that the CML Chebyshev maps have unfolded all
their chaotic state space structure, we have chosen in the simulations to use the results
produced after  $n=5000$ iteration time steps. This is a safe choice because,
normally, the CML sequences achieve their typical long-term behavior
after about 20 iteration time steps.
Averages over time steps $n$ were not performed. Rather, our aim was
to analyze a given snapshot of symbols generated by the CML system
that spatially had comparable length to that of the genomic sequence.

\par To locate the coupling constant which best reproduces the complexity of the
genomic sequences, we calculated the multifractal spectra of the genomic sequence
and compared it with the spectra of the CML-symbol sequences for various values of $\alpha$.
The estimation of the multifractal exponents $D_q$ are based on the calculation of
the probabilities $p(i_1,i_2,...i_N)$ of finding blocks of symbols $i_1,i_2,...i_N$
along the sequence of size $L$, whereas $N$ is the linear size of the block
\cite{provata:2010,beck:2010}:
\begin{eqnarray}
D_q=\frac{1}{q-1}\lim_{N \to \infty} \frac{\log {\displaystyle \sum_{i_1,i_2,...i_N} [p(i_1,i_2,...i_N)]^q}}
{\log E},\>\> q\ne 1 \\
\nonumber
D_1=\lim_{N \to \infty} \frac{\log {\displaystyle \sum_{i_1,i_2,...i_N}}p(i_1,i_2,...i_N)\ln [p(i_1,i_2,...i_N)]}
{\log E} ,
\label{eq0511}
\end{eqnarray}
In Eq. (\ref{eq0511}) $E $ represents the size (total number of configurations) of
the statespace. In this representation
the exponents $D_q$ represent the increase of the phase space when the size of the sequence
(or window) increases. As an example we consider the homogeneous case, where
\begin{eqnarray}
\left.
\begin{array}{l}
p(i_1,i_2,...i_N)={1}/{s^N}\\
\nonumber
E =s^N \\
\nonumber
s=\>\> {\rm the \>\> number\>\> of \>\> symbols}=4
\end{array}
\right\}
\Rightarrow D_q=1
\label{eq0512}
\end{eqnarray}
as is expected for homogeneous sequences.

\par
\begin{figure}
\includegraphics[clip,width=0.45\textwidth,angle=0]{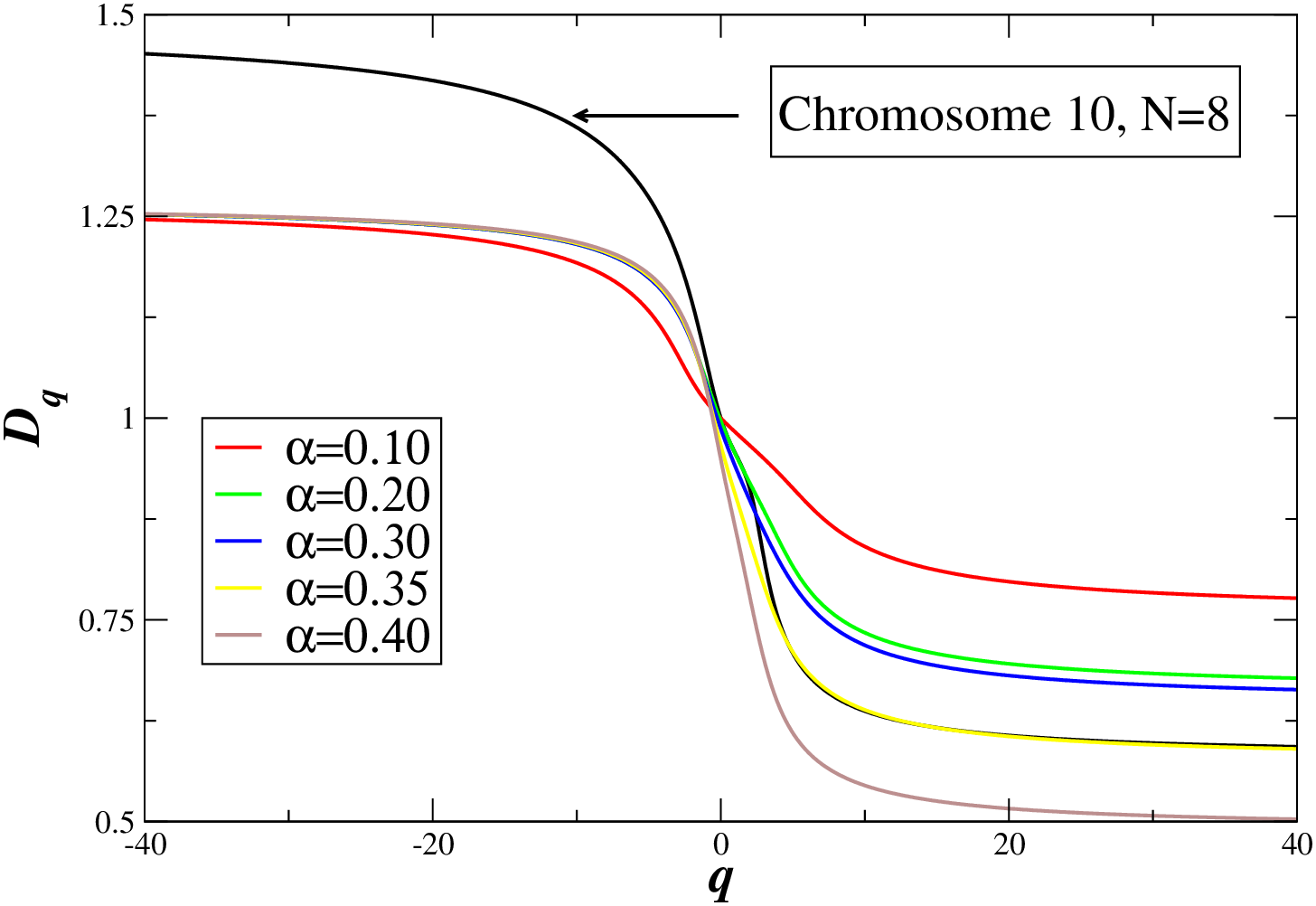}

\caption{\label{fig:03}{} (Color online)
The multifractal spectra for human Chromosome 10, with $N=8$ (black line) and
CML-$T_4$'s for various values of the coupling constant $\alpha$.}

\end{figure}

 In Fig. \ref{fig:03} the multifractal spectrum of the human Chromosome 10
is plotted and compared with
CML-$T_4$ for various values of $\alpha$.
The calculations of the spectrum of chromosome 10 are based on evaluating all possible
symbol sequences up to length $N=8$. This is considered as
asymptotic behavior since the numerical
result does not change for values $N>6$, as was previously shown in references
\cite{yu:2001,provata:2010}. For the calculation of the various spectra
based on the $T_4$ map the 1-point probabilities of the basepairs in
chromosome 10 have been respected, via Eqs. (\ref{eq04}) and (\ref{eq05}),
by choosing appropriate borders $x_1,x_2,x_3$ of the intervals.
The observed 1-point probabilities for chromosome 10,  used in this study,
are \cite{provata:2010}

\begin{eqnarray}
p_A=0.291921, \> p_C=0.207966, \\
\nonumber
p_G=0.207859, \> p_T=0.292219.
\label{eq051}
\end{eqnarray}

\par A first look at Fig. \ref{fig:03}, in the negative $q$ region,
indicates that the coarse graining of the state space into 4 segments and
the reduction of the continuous CML dynamics to a 4-symbol shift modifies
the $T_4$ spectrum, producing $D_q>1$ for $q<0$. This is not surprising, since
the multifractal spectrum presented in Eq. (\ref{eq01}) relates to the
statistics of the map and represents the frequency of iterates
within the interval [-1,1], while Eq. (\ref{eq0512}) relates to the
increase in the number of configurations in the symbolic sequence resulting
from this map.
\par
\begin{figure}
\includegraphics[clip,width=0.45\textwidth,angle=0]{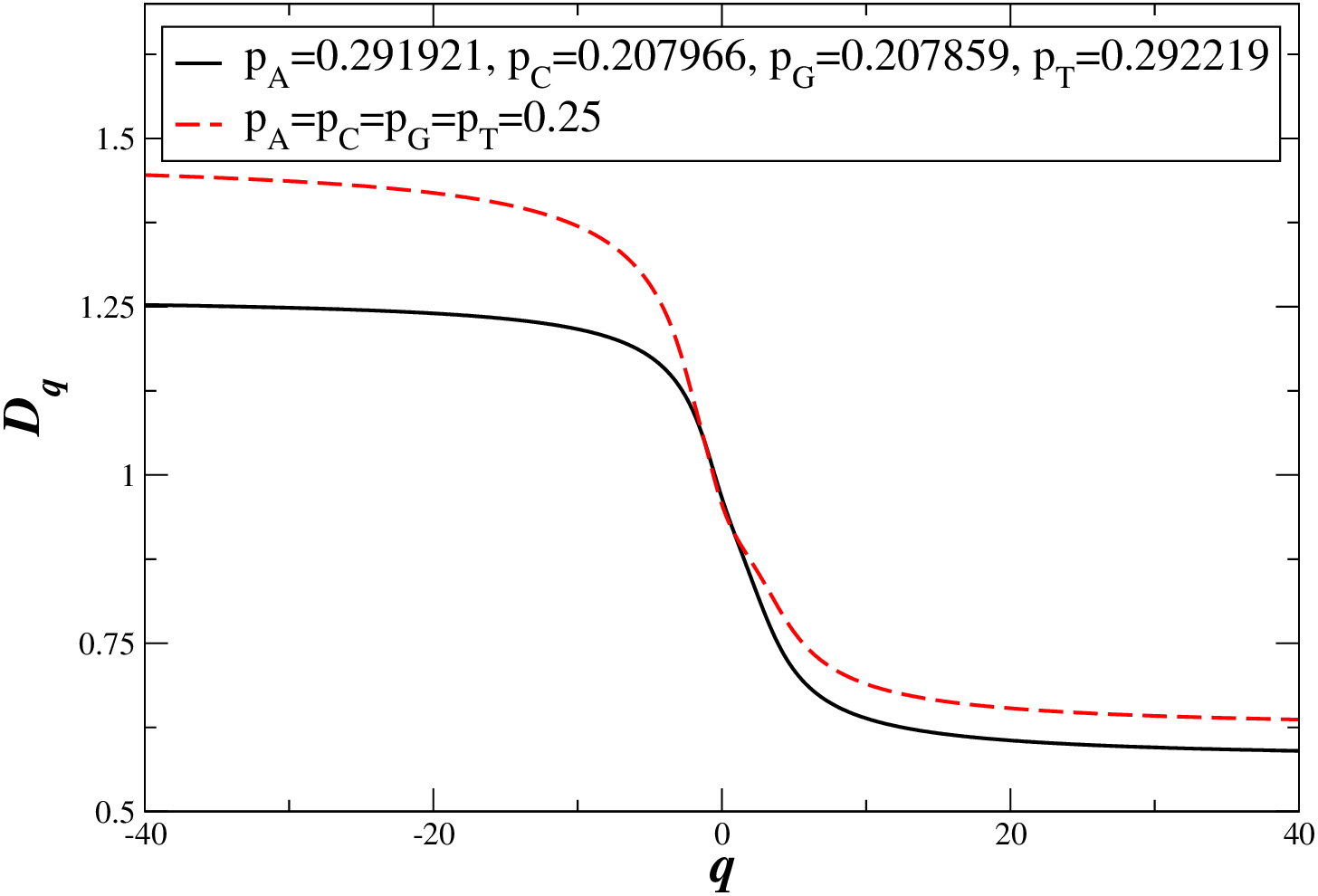}

\caption{\label{fig:04}{} (Color online)
The multifractal spectra for two artificial DNA sequences, produced via the
$CML-T_4$. The coupling constant is $\alpha =0.35$. The dashed (red) line represents
equal composition of the four basepairs $p_A=p_C=p_G=p_T=0.25$, while the solid (black) line
represents composition given by Eq. (9).}

\end{figure}

 In addition, the deviation from unity for the negative $q$ spectrum is
accentuated by the unequal frequencies of the four basepairs.
  In Fig. \ref{fig:04} the multifractal spectrum arising
from the CML assimilating a random artificial DNA sequence with equal
basepair distribution $p_A=p_C=p_G=p_T=0.25$ is compared to the case of non-equal
basepair composition, Eq. (9). In the case of equal frequencies the
CML process seems to create some rare configurations with small probability
which tend to increase the $D_q$-values in the negative $q$ region.

\par Returning to Fig.~\ref{fig:03} one observes that
the closest approximation to the Chromosome 10 spectrum is achieved for
$\alpha\sim 0.35$. This approximation is good only for positive $q-$values, which
correspond to positive moments of the probability density.
This means that configurations which are found often in the genome
are well represented by the CML process $\alpha =0.35$, while rare configurations,
which dominate the negative-$q$ spectra, are not
accounted sufficiently in this approach.

\par In the next section, we will improve the model
by {\it ad hoc} introducing a small number of rare configurations.

\section{Taking into account Rare Oligonucleotides}
\label{sec4}
\par To motivate the need for including rare oligonucleotide configurations,
we first discuss  some DNA functional issues related to the presence of specific
rare segments.
Long-time studies of the primary structure of higher eucaryotes
and other organisms have revealed certain particularities related mainly
to the functionality of DNA sequences. In particular, in coding sequences
all combinations of the four letters are found with almost equal probability, without
priority given to specific combinations. This is not true for the noncoding
parts which comprise 95-97\% of the human genome. In noncoding sequence
repetitive elements are very common with only the Alu-(repetitive sequence)
covering approximately 11\% of the human genome \cite{alu:2008}.
Other common elements which are often met in eucaryotes
are the poly-A and poly-T chains. Likewise, sequences with specific
functionality are very rare and they are only present for specific purposes
in the noncoding region. Well known such examples are the TATAA box and the $GC$ and $CG$
complexes and multiple superpositions of them
\cite{li:2002,bernaola:2004,li:2004,cheng:2005,messer:2006,katsaloulis:2002,katsaloulis:2006,han:2008,freudenberg:2009}.
The presence of these complexes is associated with the presence of promoters,
regulatory elements which designate the subsequent appearance of  coding
segments. These regulatory elements have the very specific task of ``chemically
attracting'' the enzymes which will act on the closely  following coding sequence
in order to start the production of RNA which will finally lead to the production
of the corresponding protein. Thus the presence of promoters (and their sequence
structure) is very specific in the noncoding sequences and they are not abundantly found in the genome.
Promoters are not the only sequences which are conserved for specific purposes. Other
regulatory elements, such as the cis-acting and trans-acting elements,
also have rare sequence structure.
\par From the above discussion it becomes clear that the
structure of the noncoding segments, which dominate the genome of higher eucaryotes,
is far from being uniform. The presence of rare configurations, which is mostly
visible in the negative $q$ spectrum, needs to be taken into account for a
proper modelling of the sequence dynamics. Rare sequences with
specificity, which are not accounted for by the simple CML model presented
in the previous section, will be thus considered {\it ad-hoc}
in this section. This addition will
mostly contribute to  the negative $q-$ values of the multifractal spectrum,
which is observed to be lower for the CML than for the
human chromosomes.
\par
\begin{figure}
\includegraphics[clip,width=0.45\textwidth,angle=0]{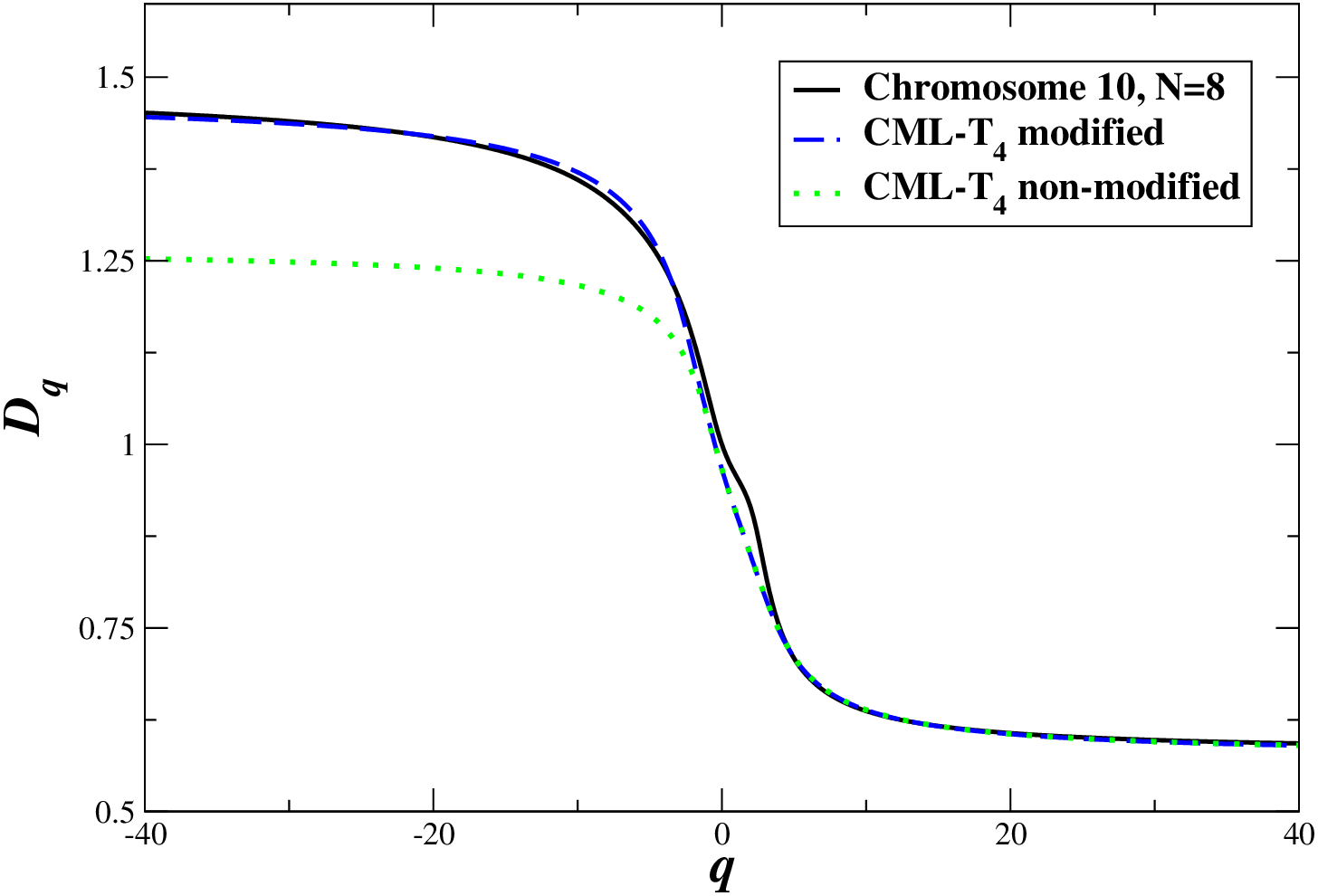}

\caption{\label{fig:05}{} (Color online)
 The multifractal spectrum for human chromosome 10, with $N=8$ (black solid line),
and CML-$T_4$ with coupling constant $\alpha =0.35$
 modified with few rare configurations as described in the text (blue dashed line).
The green dotted line depicts the CML-$T_4$ representation without rare configurations.}
\end{figure}

We modify the dynamics by assuming that some of the rare symbol sequences
generated by the CML become even less frequent by an
external coupling mechanism (such as escape from the
chaotic attractor).
It is sufficient to artificially modify the probability of
occurrence of a small fraction  $\theta $ of configurations, eg.
  $\theta \sim 1/1000$, creating thus $\Theta = \theta \times E$
rarified configurations.
The probability of occurrence of these rare configurations
can be reduced to as much as $10^{-1}\times ({\rm lowest-probability})$ for a much better
approximation of the chromosomal multifractal spectra. These values are indicative
and they depend weakly on the specific chromosome.
In Fig. \ref{fig:05} the multifractal spectrum of human chromosome 10
is plotted together with CML-$T_4$ with coupling constant $\alpha =0.35$
and with modifications to include  $\Theta =60$ rare configurations (blue line).
For comparison the case of CML-$T_4$ with coupling constant $\alpha =0.35$
without rare configurations is also included (green dotted line).
Because the modified configurations are few and rare, their contributions to the
positive $q$ region of the multifractal spectrum is negligible. On the contrary,
for negative $q$ they give an important contribution increasing significantly the
$D_q$ values and good agreement is then achieved for both positive and negative
$q$. Similar results are also obtained for the other human chromosomes, with
adjustable values of $\alpha$ and $\theta$.

\par
\begin{figure}
\includegraphics[clip,width=0.45\textwidth,angle=0]{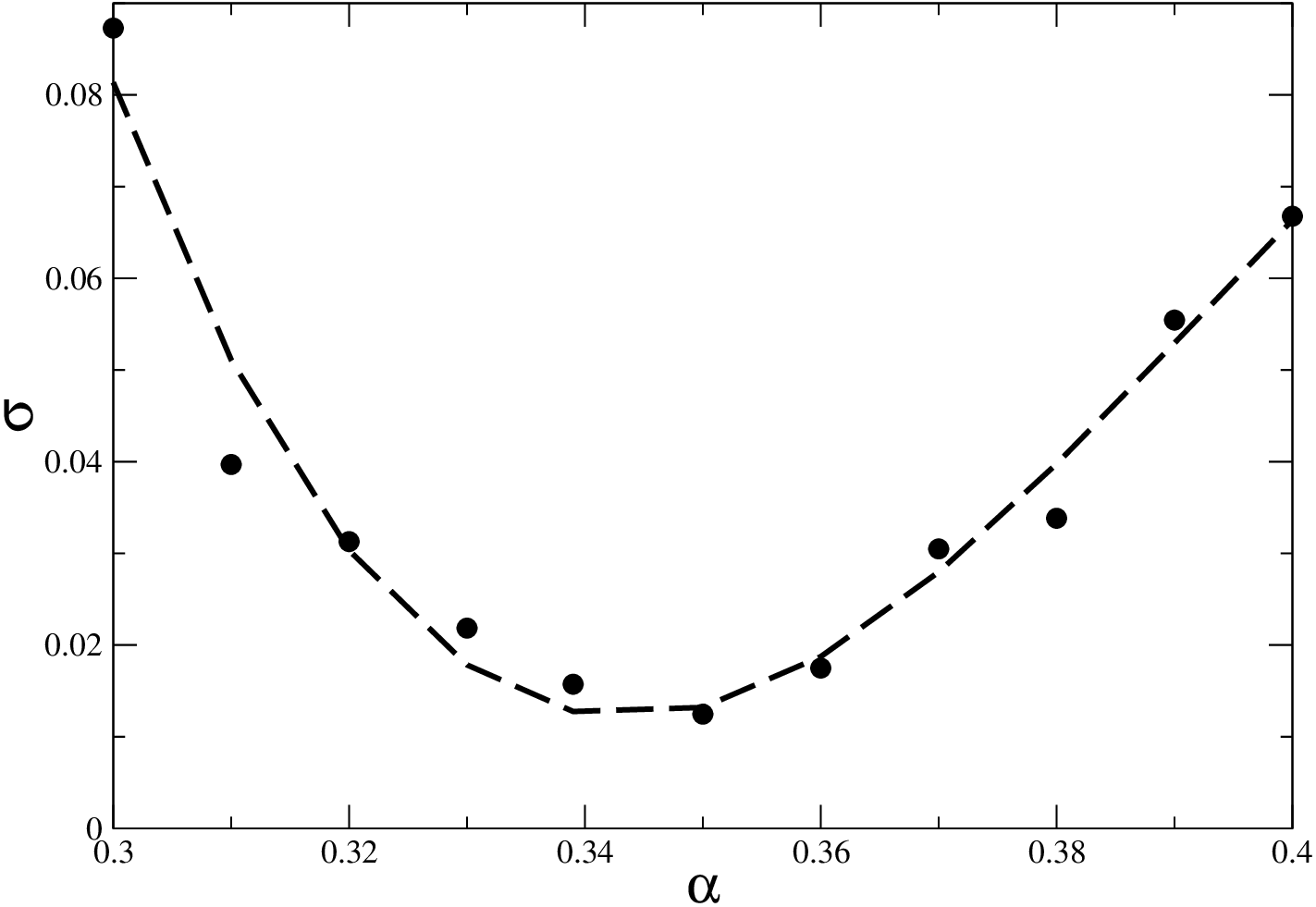}

\caption{\label{fig:07}{}
 The deviation $\sigma$ (black circles) as a function of $\alpha$ between the multifractal
spectra of human chromosome 10 and CML-$T_4$'s modified with rare configurations.
The dotted line is a cubic fit to the data.}
\end{figure}

 In Fig. \ref{fig:07} we plot the variable $\sigma$ which denotes the
mean square deviation between the multifractal curve of chromosome 10 and CML-$T_4$'s modified
with rare configurations for various values of $\alpha$.
\begin{eqnarray}
\sigma ^2(\alpha )=
\frac{1}{N_1}\sum_{q=-N}^N\left( D_q^{CML}(\alpha )
-D_q^{10}\right) ^2
\label{eq06}
\end{eqnarray}
In Eq. (\ref{eq06})
 $D_q^{CML} (\alpha )$ denotes the multifractal exponent of order $q$ for the
CML of $T_4$ with coupling constant $\alpha$, while $D_q^{10}$ denotes the
corresponding multifractal exponent of order $q$ for chromosome 10. The sum runs
from $-N$ to $N$ over positive and negative $q-$values at equal distance $\Delta$;
$N_1=(2N+1)/\Delta$ is the total number of $q-$ values considered.
Figure \ref{fig:07} shows that the smallest $\sigma$ value for chromosome 10
is found for $\alpha =0.35\pm 0.01$ and thus the coupled Chebyshev string
with coupling constant  $\alpha =0.35\pm 0.01$ best represents
the correlations found in chromosome 10. Similar
values are also obtained for the other human chromosomes.

\section{Conclusion}
\label{sec5}

The multifractal spectrum of CMLs has been determined using as a working example
the 4-th order Chebyshev map $T_4$ diffusively coupled on a 1-D
lattice with periodic boundary conditions.
This choice of local map is particularly suited for our biological application
since it corresponds to a shift of information encoded in 4 symbols,
just as the DNA string encodes information using 4 nucleotides.
\par In the current study the CML was used to assimilate the correlated
structure of genomic sequences, based on a comparison of the
corresponding multifractal spectra.
It was shown that the CML of $T_4$ can reproduce quite closely the multifractal
spectrum of genomic sequences for positive $q-$values, while it deviates
significantly for negative $q$. In particular, for human chromosome 10
(complete sequence) the best approximation for the positive $q$ spectrum
is obtained for coupling constant value $\alpha =0.35$.
\par In an attempt to model both positive and negative $q-$ spectra of
chromosomes as closely as possible, we consider the differences of the frequency
representation of various functional units.
One particular property of noncoding DNA is the presence of rare
configurations (oligonucleotide sequences) which have specific functionality
serving as promoters for the production of proteins or as regulatory elements.
Such specific sequences are the TATAA-box, various $GC$-complexes and other
elements which vary for different classes of organisms.
We model these rare configurations by introducing an
additional (artificial) escape process for the CML, which modifies
the probabilities of certain rare sequences.
If rare
configurations representing these particular sequences are considered via an
{\it ad-hoc} modification of the simulated distribution in the symbolic sequences
resulting from  the CML of $T_4$, then we see that
 both negative and positive $q$ 
multifractal spectra of genomic sequences are well approximated by the CMLs.
\par It is interesting to mention here that a good representation of the
distribution of base  pair sequences  in DNA  must be a superposition of (at least)
two components. One of these components represents mostly the coding sequences
and the second one contributes to the noncoding ones. This is in line with earlier studies
of DNA sequences which have shown that the coding and noncoding
parts follow different statistics,
related to their different functionality
\cite{scaling,provata:2007,beck:2010}.
\par In the current study, as a representative example, the human chromosome 10 was
investigated in detail and the optimum value of the
corresponding coupling constant of the CML was determined.
Likewise, additional studies not described here
have shown similar qualitative and quantitative
behavior for the other human chromosomes.
Further studies are required to show if the same approach can be applied
to different classes of organisms, where the ratio of lengths of coding/noncoding
sequences takes on different values. In a future study it will be very interesting
to explore the range of values of the coupling constant $\alpha$ and the
rare configuration frequency $\theta$
that may characterize the different classes of organisms.

\end{document}